\documentclass[showpacs,aps,prd,floatfix,amsmath,amssymb,nofootinbib,superscriptaddress]{revtex4-2}
\usepackage{graphicx}
\usepackage{epstopdf}
\usepackage{amssymb}
\usepackage{bbm}
\usepackage[utf8]{inputenc}
\usepackage[dvipsnames]{xcolor}
\usepackage{newunicodechar}
\newunicodechar{−}{-}
\usepackage{slashed}
\usepackage{float}
\usepackage{subfigure}
\usepackage{hyperref}
\usepackage{xcolor}
\begin{document}

\title{$CP$-odd form factor in the $HWW$ vertex}
\author{A.I. Hern\'andez-Ju\'arez}
\email{alan.hernandezjua@alumno.buap.mx}
\address{Facultad de Ciencias F\'isico Matem\'aticas, Benem\'erita Universidad Aut\'onoma de Puebla, Apartado Postal 1152, Puebla, Pue., M\'exico.}
\author{G. Tavares-Velasco}
\address{Facultad de Ciencias F\'isico Matem\'aticas, Benem\'erita Universidad Aut\'onoma de Puebla, Apartado Postal 1152, Puebla, Pue., M\'exico.}
\author{J. Mart\'inez-Ram\'on}
\address{Facultad de Ciencias F\'isico Matem\'aticas, Benem\'erita Universidad Aut\'onoma de Puebla, Apartado Postal 1152, Puebla, Pue., M\'exico.}

\begin{abstract}
 We revisit the $CP$-odd form factor in the $HWW$ vertex within the Standard Model (SM), which is induced only at the one-loop level when one of the $W$ bosons is off-shell. To the best of our knowledge, the numerical evaluation of this form factor is presented for the first time. The relevant contributions from quark loops are of order $10^{-5}$, well below the current experimental sensitivity. The phenomenological implications are studied through asymmetries in unpolarized and polarized observables in three-body decays.   
  \end{abstract}


\date{\today}

\maketitle

\section{Introduction}

The Higgs boson is the physical manifestation of the mechanism responsible for generating the masses of gauge bosons and fermions in the Standard Model \cite{Englert:1964et, Higgs:1964pj, Guralnik:1964eu}. Since its landmark discovery in 2012 at the Large Hadron Collider (LHC) \cite{CMS:2012qbp, ATLAS:2012yve}, substantial progress has been achieved in the precise determination of its properties \cite{ATLAS:2022vkf, CMS:2022dwd}. Recent measurements by the ATLAS and CMS collaborations have reported updated determinations of Higgs decays into $Z\gamma$ \cite{CMS:2022ahq, ATLAS:2023yqk} and $ZZ$ final states \cite{CMS:2022ley, ATLAS:2023dnm}. Motivated by these developments, we investigate the $HWW$ vertex, as ongoing LHC analyses may yield new insights into this interaction.

The decay $H\to WW$ within the SM has been extensively studied in the literature \cite{Lee:1977eg, Rizzo:1980gz, Fleischer:1981um, Bredenstein:2006rh, Bredenstein:2006nk, Bredenstein:2006ha, Huang:2020zde}. Experimentally, the process $H\to WW^\ast$ has been examined in both gluon-fusion and vector-boson-fusion production channels, with measurements consistent with SM expectations \cite{CMS:2013btf, ATLAS:2014aga,ATLAS:2016neq, CMS:2018zzl, ATLAS:2018xbv, ATLAS:2021pkb, ATLAS:2022ooq, CMS:2022uhn, ATLAS:2023pwa, ATLAS:2025hki}.
 One-loop corrections to the $HWW$ coupling have been evaluated for several kinematic configurations, including $H^\ast {W^{-}}^\ast {W^+}^{\ast}$ \cite{Kniehl:1991xe, KNIEHL1994211}, $H^\ast {W^{-}} {W^+}$ \cite{Phan:2023xgi}, and the fully on-shell $HW^-W^+$ vertex \cite{Fleischer:1981um}. Moreover, radiative corrections have been computed in a variety of extensions of the SM, including the SMEFT \cite{Dawson:2018liq}, heavy-scalar scenarios \cite{Ria:2017qxd}, the minimal supersymmetric SM \cite{Hollik:2011xd}, the Next-to-Minimal Supersymmetric Standard Model (NMSSM) \cite{Domingo:2018uim}, and models with Majorana neutrinos \cite{Ilakovac:1993pt}. These corrections to the $HWW$ vertex, obtained across multiple theoretical frameworks, have been incorporated into the computational package \texttt{H-COUP} \cite{Kanemura:2017gbi}. In addition, the decay of the Higgs boson into $W$ pairs has been implemented in widely used public tools such as \texttt{HDECAY} \cite{Djouadi:1997yw} and \texttt{PROPHECY4F} \cite{Bredenstein:2006rh}. Further studies of the $HWW$ coupling can be found in Refs.~\cite{Barger:1993wt, Dittmar:1996ss, Moretti:1997ng, Zerwas:1999pm, Kniehl:2000kk, Han:2000mi, Djouadi:2005gi, Dutta:2008bh, Bao:2010sz, Masso:2012eq, Gonzalez:2012mq, Banerjee:2015bla, Brehmer:2019gmn, Sharma:2022epc, Ma:2021cxg, Goncalves:2025xer, Novales-Sanchez:2025ilh, Colyer:2025ehv}.

\begin{figure}[H]
\begin{center}
\includegraphics[width=9cm]{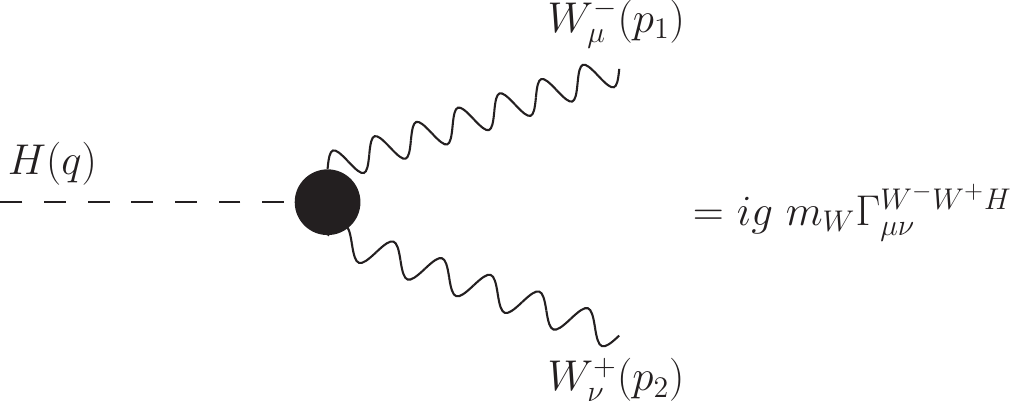}
\caption{Kinematic configuration and momentum assignments for the $HWW$ vertex. The graphics in this work are created with \texttt{JaxoDraw} \cite{Binosi:2003yf}} \label{HWWvertex}
\end{center}
\end{figure}

The $HWW$ vertex can be described by considering the kinematics of $H\rightarrow W^-W^+$ and adopting the momentum conventions shown in Fig.~\ref{HWWvertex}. The corresponding vertex function can then be written as 
\begin{align}\label{vertxeq}
\Gamma_{\mu\nu}^{W^-W^+H}(p_1^2,p_2^2,q^2)=h^V_1(q^2,p_1^2,p_2^2) g_{\mu\nu}+\frac{h_2^V(q^2,p_1^2,p_2^2)}{m_W^2} p_{1\nu}p_{2\mu}+\frac{h_3^V(q^2,p_1^2,p_2^2)}{m_W^2}\epsilon_{\mu\nu\alpha\beta}p_1^\alpha p_2^\beta,
\end{align}
where $h_i^V$ denote the form factors that parametrize the interaction. Since the three particles cannot be simultaneously on-shell, we focus on configurations in which one of the bosons is virtual, with $V$ labeling the off-shell leg ($V=H$, $W^-$, or $W^+$). In terms of the anomalous couplings, the general expressions for the form factors are given by \cite{Hernandez-Juarez:2023dor}
\begin{align}
& h_1^V(q^2,p_1^2,p_2^2)=1+ a_W - \hat{b}_W \frac{q^2-p_1^2-p_2^2}{m_W^2}+ \frac{\hat{c}_W}{2} \frac{p_1^2+p_2^2}{m_W^2},\\
&h_2^V(q^2,p_1^2,p_2^2)= 2 \hat{b}_W,\\
&h_3^V(q^2,p_1^2,p_2^2)= 2 \widetilde{b}_W.
\end{align}
The form factors $h^V_{1,2}$ are CP-conserving, whereas $h_3^V$ is $CP$-odd. In the SM, $h_1^V=1$ at tree level, while the remaining form factors vanish. At the one-loop level, the anomalous couplings $\hat{b}_W$ and $\hat{c}_W$ are generated, with $\hat{b}_W$ being complex and of order $10^{-2}$-$10^{-3}$. The CP-odd form factor $h_3^V$ also arises at one loop, but only when a $W^\pm$ boson is off-shell \cite{Kniehl:1991xe, KNIEHL1994211}. This form factor has been investigated in a wide range of theoretical scenarios \cite{Fleischer:1981um, Soni:1993jc, Ilakovac:1993pt, Han:2000mi, Montano:2005kt, Takubo:2010tc, Desai:2011yj, Biswal:2012mp, Anderson:2013afp, Godbole:2013saa, Dwivedi:2015nta, Ferreira:2016jea, Nakamura:2017ihk, Rao:2018abz, Huang:2020zde, Rossia:2024rfo, Das:2025ifh, Silva:2025hzo}, as well as in studies of its phenomenological impact at the LHC and future colliders \cite{Kniehl:2001jy, Biswal:2005fh, Choudhury:2006xe, Biswal:2008tg, Biswal:2009ar}. The $CP$-odd form factor in the $HWW$ coupling can induce novel observables, as seen in the analogous $HZZ$ vertex \cite{Godbole:2007cn}. 

In this study, we revisit $h_3^V$ in the $HWW$ vertex within the SM. The structure of this work is organized as follows. In Sec. \ref{CPsm}, we study the one-loop contribution to the $CP$-odd form factor $h_3^V$ in the $HWW$ coupling. Furthermore, we present the first numerical evaluation of this contribution. In Sec. \ref{sec3bodydec}, we analyze the implications of $h_3^V$ in the $H\to \overline{f} f W$ decay, where forward-backward and left-right asymmetries are introduced. Next, in Sec. \ref{prospectsSec}, we discuss the experimental prospects for observing these asymmetries at the LHC. Finally, in Sec. \ref{concSec}, we present our conclusions. 


\section{The CP-odd form factor in the SM}\label{CPsm}

\begin{figure}[H]
\begin{center}
\includegraphics[width=8cm]{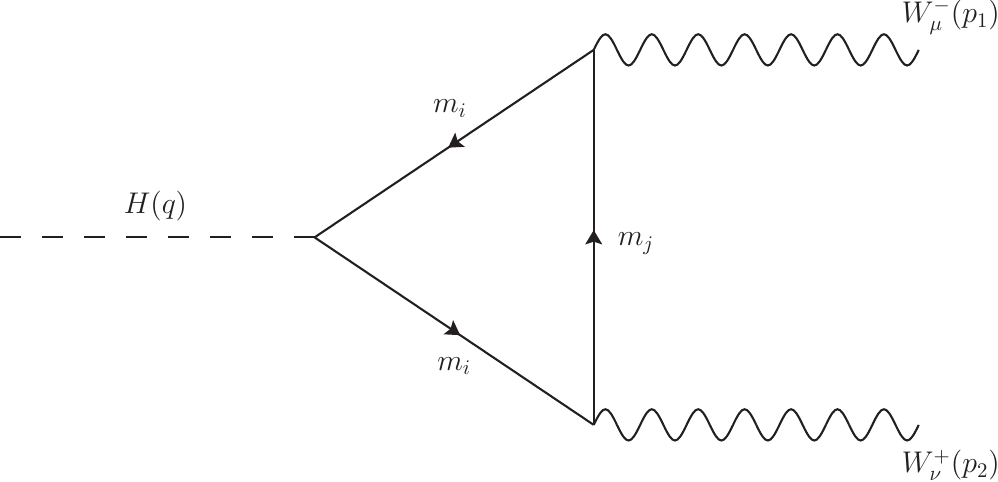}
\caption{Fermion loop contribution to the $HWW$ vertex.} \label{diag}
\end{center}
\end{figure}

In this section, we revisit the SM contribution to the CP-odd form factor at the one-loop level. This result was first reported in Ref.~\cite{Kniehl:1991xe}. In the $HWW$ vertex, $h_3^V$ arises from the fermion-loop diagram shown in Fig.~\ref{diag}. Using the \texttt{FeynArts} \cite{Hahn:2000kx} and \texttt{FeynCalc} \cite{Mertig:1990an, Shtabovenko:2016sxi, Shtabovenko:2020gxv, Shtabovenko:2023idz} packages, we compute the general CP-odd form factor $h_3^V$, which is given as follows
\begin{align}
\label{h3}
h_3^V=&\frac{i g^2 N_c \big| V_{ij}\big|^2 m_i^2}{32 \pi ^2 
   \big[(p_1-p_2)^2-q^2\big] \big[(p_1+p_2)^2-q^2\big]}
   \Big\{   \big[p_1^2-p_2^2\big]
 \big[\big(2 m_i^2-2
   m_j^2+p_1^2+p_2^2-q^2\big)
   \text{C}_0\left(p_1^2,p_2^2,q^2,m_i^2
   ,m_j^2,m_i^2\right)\nonumber\\
   &-2
   \text{B}_0\left(q^2,m_i^2,m_i^2\right)\big]+ \left(3
   p_1^2+p_2^2-q^2\right)
   \text{B}_0\left(p_1^2,m_i^2,m_j^2\right)-
   \left(p_1^2+3 p_2^2-q^2\right)
   \text{B}_0\left(p_2^2,m_i^2,m_j^2\right)-(m_i\leftrightarrow m_j)\Big\}
\end{align}
where $\big|V_{ij}\big|$ denotes the norm of the $ij$ element of the CKM matrix, which for leptons is $\big|V_{ij}\big|=1$, and $N_c$ corresponds to the number of colors. The different $\text{B}_0$ functions are divergent. However, they contain the same divergence (see Appendix~\ref{b0ap}), which cancels analytically against the $(m_i\leftrightarrow m_j)$ contribution due to the opposite sign of identical terms. Furthermore, the amplitude is gauge independent because the Feynman diagram in Fig.~\ref{diag} contains only fermion propagators. As a consequence, the $h_3^V$ form factor does not depend on the gauge parameter, even when one of the $W$ bosons is off shell. This situation differs from that of the $CP$-even form factors, where gauge-dependent contributions  appear. In such cases, gauge independence can be consistently implemented through the pinch technique \cite{Cornwall:1981zr, Cornwall:1989gv, Binosi:2009qm} or the background field method \cite{Denner:1994xt, Pilaftsis:1996fh}.

From Eq. \eqref{h3}, it follows that in the SM the $CP$-odd form factor is induced only when the $W$ bosons have different virtualities ($p_1^2 \neq p_2^2$), whereas it vanishes for on-shell $W$ bosons. Moreover, $h_3^V$ is antisymmetric under the exchange $p_1 \leftrightarrow p_2$. As a consequence, in the decay $H\to W^*W$ the form factor changes sign depending on whether the $W^-$ or $W^+$ boson is off-shell. The expression in Eq.~\eqref{h3} is consistent with that reported in Ref.~\cite{Kniehl:1991xe}. 

In the SM, $CP$-violating effects are characterized by the Jarlskog invariant $\mathcal{J}$ \cite{Wu:1985ea, Jarlskog:1985ht, Huang:2020zde, Luo:2023fcc, Luo:2025wio}, defined as
\begin{equation}
{\rm Im}\big(V_{ij}V_{kl}V^\ast_{il} V^\ast_{kj} \big)=\mathcal{J}\sum_{m=1}^3\epsilon _{ikm}\sum_{n=1}^3\epsilon _{jln}.
\end{equation}
Following the PDG parametrization \cite{ParticleDataGroup:2026aaa}, the invariant is given by 
\begin{equation}
\mathcal{J}=c_{12}s_{12}c_{13}^2s_{13}c_{23}s_{23}\sin\delta,
\end{equation}
where $c_{ij}=\cos\theta_{ij}$ and $s_{ij}=\sin\theta_{ij}$, with $\theta_{ij}$ denoting the quark-family mixing angles and $\delta$ the complex phase of the CKM matrix. It is noted that the form factor $h_3^V$ is not proportional to $\mathcal{J}$ and therefore does not represent a genuine source of $CP$ violation. In fact, the CKM matrix is not required for the generation of $h_3^V$, as analogous contributions also arise from lepton loops. Instead, this $CP$-odd form factor is induced by absorptive phases in the loop integrals, which originate from on-shell intermediate states and can be computed using the Cutkosky cutting rules \cite{Cutkosky:1960sp}. 

\begin{figure}[!htb]
\begin{center}
\subfigure{}\includegraphics[width=9.05cm]{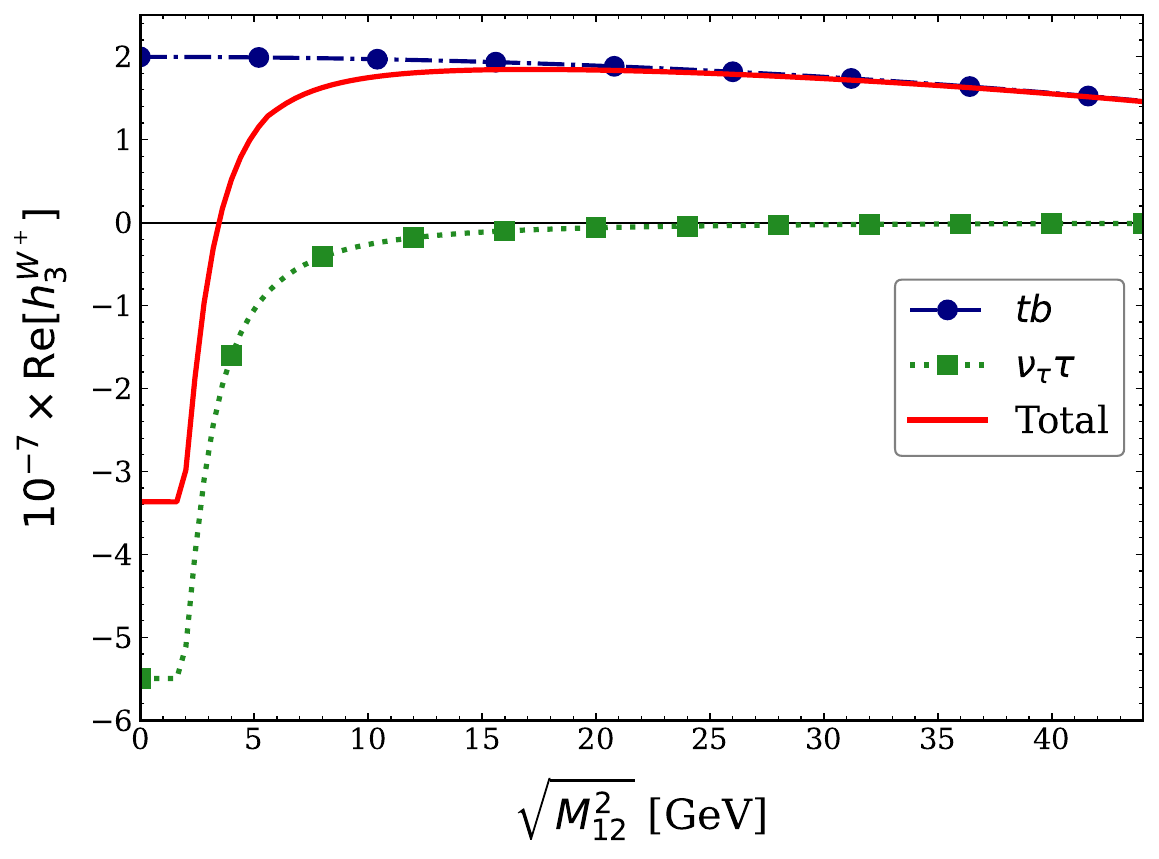}\hspace{-.32cm}
\subfigure{}\includegraphics[width=9.05cm]{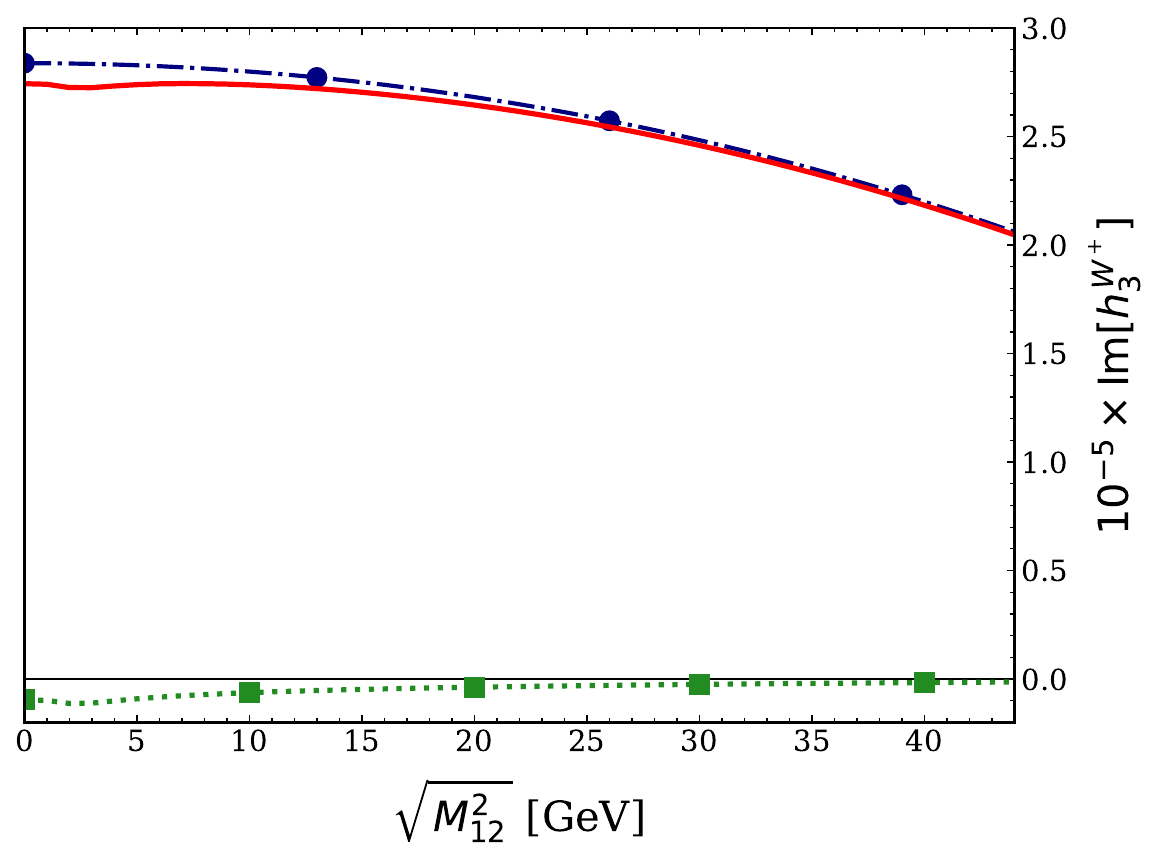}
\caption{Relevant SM contributions to the $CP$-odd form factor $h^{W^+}_3$. This scenario corresponds to the decay $H\to W^{-}W^{+*}$. For the case $h^{W^-}_3$, the values are identical but change sign.  \label{Ploth3}}
\end{center}
\end{figure}

To the best of our knowledge, a numerical evaluation of the contributions to the form factors $h_3^{W^\pm}$ has not been presented before. In Fig. \ref{Ploth3}, we show the individual and total contributions to the CP-odd form factor $h_3^{W^{+}}$ as a function of the invariant mass ($M_{12}^2$) of the off-shell $W^+$ boson. The dominant contributions arise from loops involving $tb$ quarks and $\nu_\tau \tau$ leptons. For the numerical evaluation, we use $m_b = 4.18$ GeV, as recommended by the LHCHWG \cite{LHCHiggsCrossSectionWorkingGroup:2016ypw}. Since the particles coupled to the Higgs can go on-shell, an imaginary part of $h_3^{W^+}$ is generated. This absorptive contribution reaches values of order $10^{-5}$, with the $tb$ loop providing the largest effect. The real part is smaller by roughly two orders of magnitude and is instead dominated by the lepton loop at low $M_{12}^2$, whereas at higher energies the contribution from the $tb$ loop becomes the leading one. 

A similar pattern, where the dominant contribution corresponds to the imaginary part, has been observed in quark-gluon coupling \cite{Hernandez-Juarez:2020drn,Hernandez-Juarez:2020gxp, Hernandez-Juarez:2021xhy}, in trilinear gauge-boson couplings \cite{Gounaris:2000tb, Hernandez-Juarez:2021mhi, Hernandez-Juarez:2022kjx}, and in the $HZZ$ and $HZ\gamma$ vertices \cite{Hernandez-Juarez:2023dor, Hernandez-Juarez:2024zpk, Hernandez-Juarez:2025nzh}. Due to the antisymmetry of Eq.~\eqref{h3} under $p_1$ and $p_2$, the values obtained for the case with an off-shell $W^-$ boson are identical in magnitude but opposite in sign.

The current bound on the CP-odd form factor has been obtained by the ATLAS collaboration through $H\to WW^\ast \to 2\ell 2\nu$ decays in gluon-gluon and vector-boson fusion \cite{ATLAS:2025hki}. The result is expressed in terms of the SMEFT Wilson coefficient $C_{H\widetilde{W}}$ \cite{Brivio:2017vri}, which is constrained as
 \begin{equation}
\label{limcp}
\big| C_{H\widetilde{W}}\big|\lesssim 0.9.
\end{equation}
This bound represents an improvement of approximately 50\% relative to previous limits \cite{ATLAS:2023pwa}. The relation between $C_{H\widetilde{W}}$ and the anomalous coupling $\widetilde{b}_W$ in Eq.~\eqref{vertxeq} is given by $\widetilde{b}_W = 4/g^2\ C_{H\widetilde{W}}$ \cite{Barducci:2025ati}, which implies bounds of order $10^{1}$ for $h_3^{W^\pm}$. This value is five orders of magnitude larger than the SM prediction.

Although $h_3^{W^\pm}$ is not a source of $CP$ violation, it nevertheless represents a genuine one-loop prediction of the SM. The absorptive phase associated with $h_3^{W^\pm}$ is particularly relevant because, if an additional amplitude with a weak phase contributes to the same process, their interference can generate observable $CP$-violating asymmetries \cite{Branco:1999fs, Chen:2014ona}. Furthermore, in searches for observables sensitive to $CP$ violation in the $HWW$ coupling, the contribution from $h_3^{W^\pm}$ constitutes an irreducible SM background that must be properly accounted for in both phenomenological predictions and experimental analyses. This motivates a detailed study of the effects of $h_3^{W^\pm}$ in the three-body decay $H\to W^\ast W\to \overline{f}_i f_j W$.

\section{The $H\to \overline{f} f W$  decay and $CP$-odd effects}\label{sec3bodydec}

In this section, we investigate the impact of the CP-odd form factor $h_3^{W^\pm}$ on the three-body decay $H\to W^\ast W\to \overline{f}_i f_j W$. We treat $h_{1,2,3}^{W^\pm}$ as complex quantities:
\begin{equation}
h_{i}^{W^\pm}={\rm Re}\big[h_i^{W^\pm}\big]+i{\rm Im}\big[h_i^{W^\pm}\big]\text{,}\quad i=1\text{, }2 \text{, } 3.
\end{equation}
 Together with the $CP$-odd form factors, the absorptive contributions give rise to effects that are particularly relevant in polarization-sensitive observables \cite{Ilakovac:1993pt}. Therefore, we compute polarized amplitudes following the approach developed in Refs.~\cite{Hernandez-Juarez:2023dor,Hernandez-Juarez:2024zpk}. The polarization states of the $W$ bosons in Higgs decays have been studied in Refs.~\cite{Chang:1992tu, Ilakovac:1993pt, Hagiwara:1993qt, Nakamura:2017ihk, Rao:2018abz, Colyer:2025ehv}. 

\subsection{Polarized amplitudes}

\begin{figure}[!htb]
\begin{center}
\includegraphics[width=8cm]{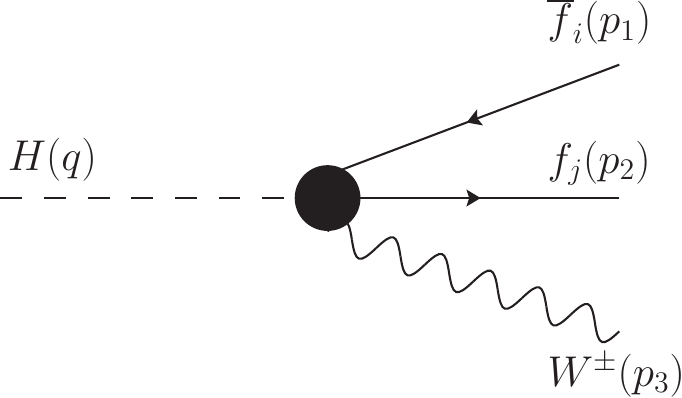}
\caption{Momentum configuration of the $H\to \overline{f}_i f_j W$ decay.} \label{diag3body}
\end{center}
\end{figure}

We use the momentum assignment shown in Fig.~\ref{diag3body} and the kinematics described in Appendix~\ref{kin1}. For a three-body decay, the total double-differential partial width can be decomposed into polarized contributions as follows:
\begin{align}
\label{doubdifEq}
\frac{d\Gamma(H\to \overline{f}_i f_j W)}{dM^2_{23}dM^2_{12}}&=\sum_\lambda\frac{d\Gamma^\lambda(H\to \overline{f}_i f_j W_\lambda)}{dM^2_{23}dM^2_{12}}\text{,}\quad \lambda=L\text{, } R\text{, }0,
\end{align}
where $M_{23}$ and $M_{12}$ are the invariant masses of the $f_j(p_2) W(p_3)$ and $\overline{f}_i(p_1) f_j(p_2)$ systems, respectively. The polarization state of the on-shell $W$ boson is denoted by $\lambda$. According to Eq.~\eqref{dif}, the invariant mass $M_{23}^2$ can be replaced by $\cos\theta$ as an integration variable, where $\theta$ is defined as the angle between the $x$ axis and $\overline{f}_i$ in the $W^{\ast}$ rest frame (see Fig.~\ref{plano}). Therefore, the double-differential partial width can be written as
\begin{align}
\label{doubdifEq2}
\frac{d\Gamma^\lambda(H\to \overline{f}_i f_j W)}{dM^2_{23}dM^2_{12}}&=-\frac{1}{2} \sqrt{(m_H^2 - m_W^2)^2 - 2(m_H^2 + m_W^2) M_{12}^2 + M_{12}^4}\frac{d\Gamma^\lambda(H\to \overline{f}_i f_j W)}{d\cos{\theta}dM^2_{12}}\nonumber\\
&= \frac{g^4  N_c \big| V_{ij}\big|^2\sqrt{(m_H^2 - m_W^2)^2 - 2(m_H^2 + m_W^2) M_{12}^2 + M_{12}^4}}{32\cdot3072  \pi^3 m_H^3 m_W^2 (m_W^2-M_{12}^2)^2}\mathcal{M}^2_\lambda(\cos\theta \text{, } M_{12}^2),
\end{align}
where the polarized amplitudes $\mathcal{M}^2_\lambda(\cos\theta \text{, } M_{12}^2)$ are given by
  
  \begin{align}\label{dpolL}
   \mathcal{M}^2_{L/R}(\cos\theta \text{, } M_{12}^2)= &\mathcal{H}(\cos\theta) M_{12}^2 \Big[
4 \Big({\rm Re}\big[h_1^{V}\big]^2+ {\rm Im}\big[h_1^{V}\big]^2  \Big) m_W^4
+ \Big({\rm Re}\big[h_3^{V}\big]^2+{\rm Im}\big[h_3^{V}\big]^2\Big) (m_H^2 - m_W^2)^2
\nonumber\\
&- 2 \Big( {\rm Re}\big[h_3^{V}\big]^2+{\rm Im}\big[h_3^{V}\big]^2  \Big) (m_H^2 + m_W^2) M_{12}^2
+ \Big( {\rm Re}\big[h_3^{V}\big]^2+{\rm Im}\big[h_3^{V}\big]^2 \Big) M_{12}^4 \nonumber\\&
\pm 4 m_W^2
\sqrt{(m_H^2 - m_W^2)^2 - 2(m_H^2 + m_W^2) M_{12}^2 + M_{12}^4}\Big( {\rm Re}\big[h_1^{V}\big]\,{\rm Im}\big[h_3^{V}\big]
- {\rm Im}\big[h_1^{V}\big]\,{\rm Re}\big[h_3^{V}\big] \Big) 
\Big],
\end{align}
with the angular function $\mathcal{H}(\cos\theta)$ defined as

\begin{align}
\mathcal{H}(\cos\theta)=\left\{
\begin{array}{l}
(1-\cos\theta)^2, \text{ for $L$,}\\[6pt]
\frac{(1-\cos^2\theta)^2}{(1-\cos\theta)^2}, \text{ for $R$.}
\end{array}\right.\end{align}
For longitudinal polarization, the corresponding amplitude reads
 \begin{align}\label{pol0d}
   \mathcal{M}^2_0(\cos\theta \text{, } M_{12}^2)= &\frac{(1-\cos^2{\theta})}{2 m_W^2}\Big[
(m_H - m_W)^2 (m_H + m_W)^2
\Big\lbrace
\big({\rm Im}[h_2^{V}]^2 + {\rm Re}[h_2^{V}]^2\big) m_H^4
-2 \big(-2\,{\rm Im}[h_1^{V}]\,{\rm Im}[h_2^{V}] \nonumber\\
&
+{\rm Im}[h_2^{V}]^2
+{\rm Re}[h_2^{V}]\,(-2\,{\rm Re}[h_1^{V}] + {\rm Re}[h_2^{V}])\big) m_H^2 m_W^2
+\Big\lbrace(-2\,{\rm Im}[h_1^{V}] + {\rm Im}[h_2^{V}])^2\nonumber\\
&
+(-2\,{\rm Re}[h_1^{V}] + {\rm Re}[h_2^{V}])^2\Big\rbrace m_W^4
\Big\rbrace
-4 (m_H - m_W)(m_H + m_W)
\Big\lbrace
\big({\rm Im}[h_2^{V}]^2 + {\rm Re}[h_2^{V}]^2\big) m_H^4 \nonumber\\
&
+3\big({\rm Im}[h_1^{V}]\,{\rm Im}[h_2^{V}]
+{\rm Re}[h_1^{V}]\,{\rm Re}[h_2^{V}]\big) m_H^2 m_W^2
+\big((2\,{\rm Im}[h_1^{V}] - {\rm Im}[h_2^{V}])({\rm Im}[h_1^{V}] + {\rm Im}[h_2^{V}])\nonumber\\
&
+(2\,{\rm Re}[h_1^{V}] - {\rm Re}[h_2^{V}])({\rm Re}[h_1^{V}] + {\rm Re}[h_2^{V}])\big)m_W^4
\Big\rbrace M_{12}^2
+2\Big\lbrace
3\big({\rm Im}[h_2^{V}]^2 + {\rm Re}[h_2^{V}]^2\big)m_H^4 \nonumber\\
&
+2\big(3\,{\rm Im}[h_1^{V}]\,{\rm Im}[h_2^{V}]
+{\rm Im}[h_2^{V}]^2
+{\rm Re}[h_2^{V}](3\,{\rm Re}[h_1^{V}] + {\rm Re}[h_2^{V}])\big)m_H^2 m_W^2\nonumber\\
&
+\big(2\,{\rm Im}[h_1^{V}]^2 + 2\,{\rm Re}[h_1^{V}]^2
+2\,{\rm Im}[h_1^{V}]\,{\rm Im}[h_2^{V}]
+2\,{\rm Re}[h_1^{V}]\,{\rm Re}[h_2^{V}]\nonumber\\
&
+3({\rm Im}[h_2^{V}]^2+{\rm Re}[h_2^{V}]^2)\big)m_W^4
\Big\rbrace M_{12}^4
-4\Big\lbrace
({\rm Im}[h_2^{V}]^2 + {\rm Re}[h_2^{V}]^2) m_H^2
+\big({\rm Im}[h_2^{V}]({\rm Im}[h_1^{V}]\nonumber\\
&+{\rm Im}[h_2^{V}])
+{\rm Re}[h_2^{V}]({\rm Re}[h_1^{V}]+{\rm Re}[h_2^{V}])\big)m_W^2
\Big\rbrace M_{12}^6
+\big({\rm Im}[h_2^{V}]^2+{\rm Re}[h_2^{V}]^2\big) M_{12}^8
\Big],
\end{align}
where in Eqs. \eqref{dpolL} and \eqref{pol0d}, the superscript in $h_i^V$ ($i=1$, 2, 3) denotes the off-shell $W$ boson. This distinction is necessary because the amplitudes $\mathcal{M}^2_{L,R}$ depend linearly on $h_3^V$, which changes sign depending on whether the $W^+$ or $W^-$ boson is off-shell.

We observe that the longitudinally polarized  amplitude is not sensitive to $h_3^V$. This behavior has also been observed in the $HZZ$ vertex \cite{Hernandez-Juarez:2023dor}. In contrast, the transverse polarizations in Eq.~\eqref{dpolL} behave differently with respect to $\cos\theta$ and the $CP$-odd form factor, as summarized below:
\begin{itemize}
  \item The function $\mathcal{M}^2_L(\cos\theta , M^2_{12})$ is proportional to $(1-\cos\theta)^2$, whereas $\mathcal{M}^2_R(\cos\theta , M_{12}^2)\sim (1-\cos^2\theta)/(1-\cos\theta)^2$.
  \item The interference terms involving the real and imaginary parts of $h_1^{V}$ and $h_3^{V}$ differ by an overall minus sign between the two transverse polarizations.
\end{itemize}

\begin{figure}[!htb]
\begin{center}
\subfigure{}\includegraphics[width=9.05cm]{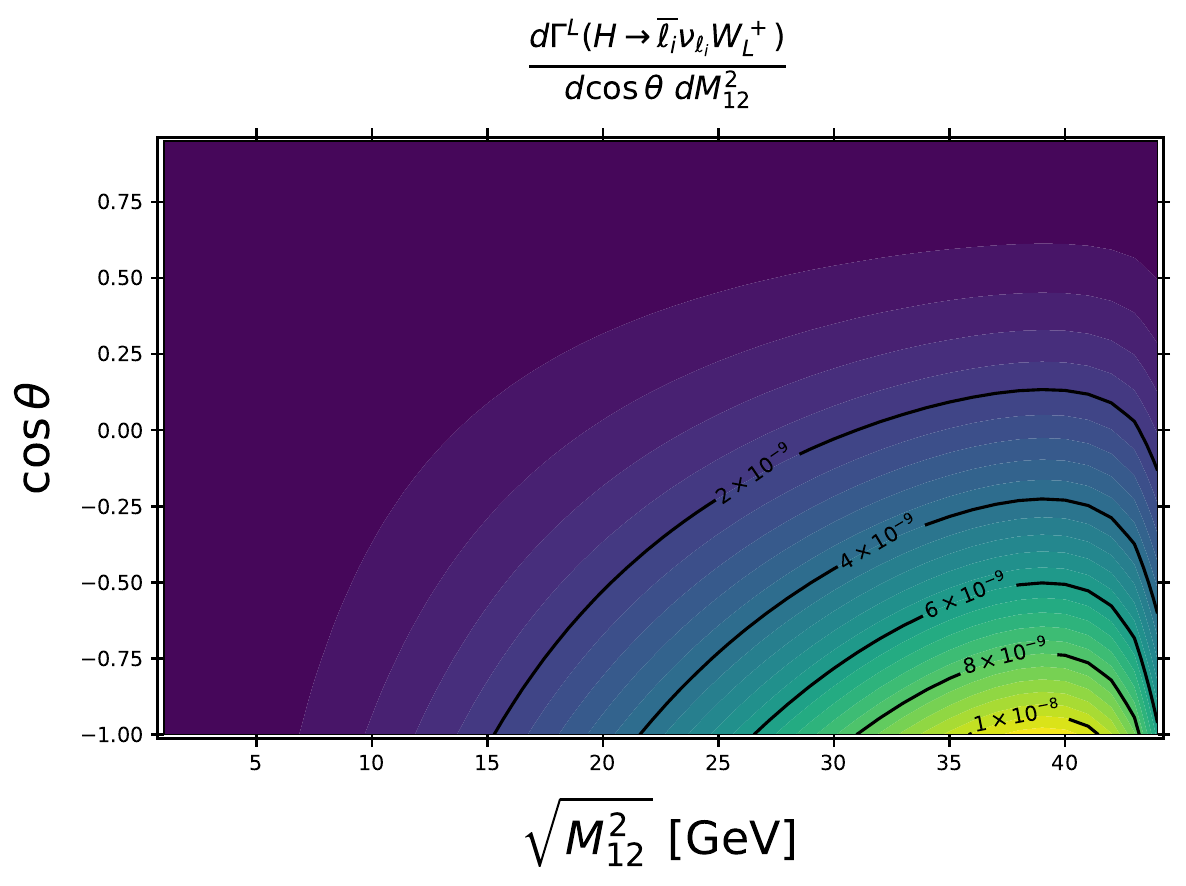}\hspace{-.28cm}
\subfigure{}\includegraphics[width=9.05cm]{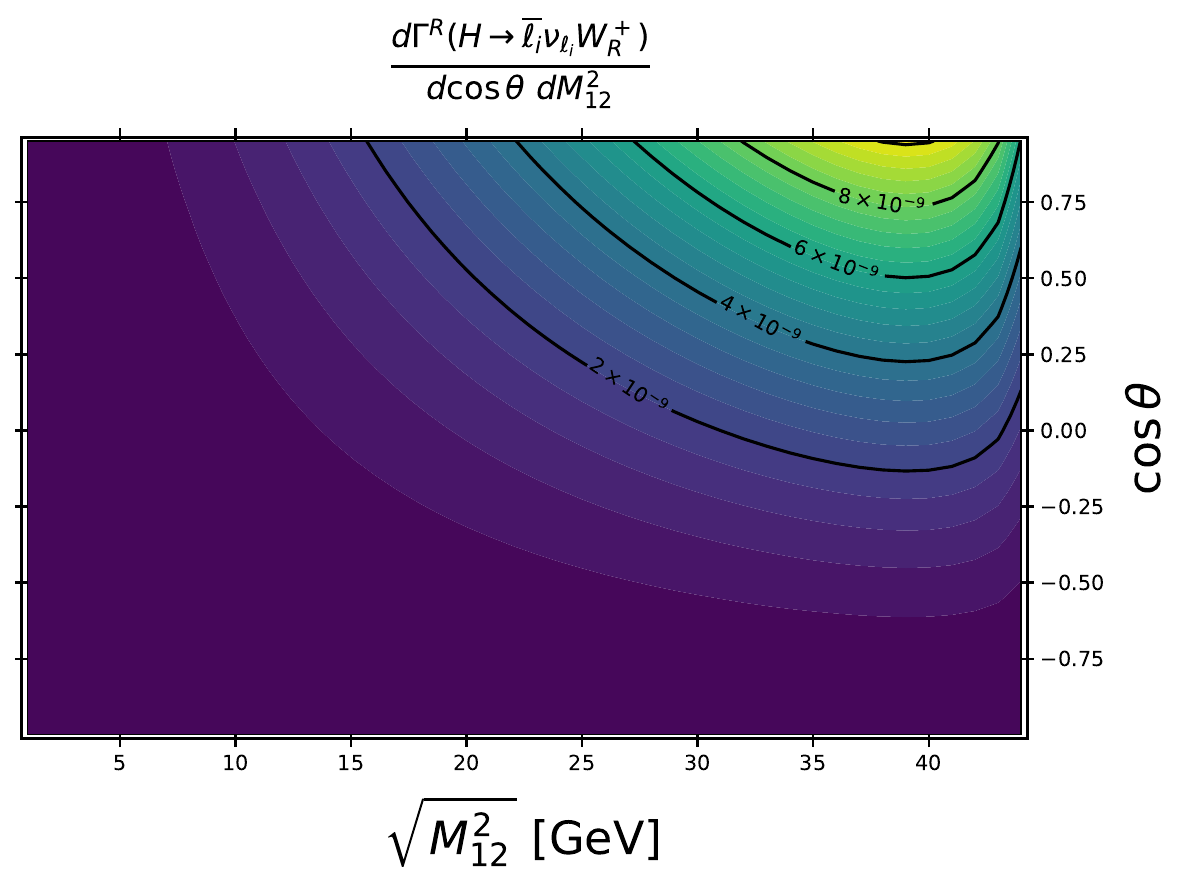}
\caption{Left- and right-polarized partial widths in the $\sqrt{M_{12}^2}$ versus $\cos\theta$ plane. The curves are obtained using the SM one-loop contributions to the form factors $h_1^{W^-}$ and $h_3^{W^-}$.  \label{Plotddif}}
\end{center}
\end{figure}

 These differences indicate distinct patterns in the polarized partial widths of the $H\to\overline{f}_i f_j W$ decay, which may give rise to phenomenologically relevant effects.  Fig.~\ref{Plotddif} shows the contours of the SM double-differential partial widths associated with transverse polarizations in the $\sqrt{M_{12}^2}$ versus $\cos\theta$ plane. We consider the scenario in which the $W^-$ boson is off-shell, with leptons in the final state. Since Eq.~\eqref{dpolL} also depends on the form factor $h_1^{W^-}$, we compute its SM one-loop contributions using the \texttt{FeynArts} \cite{Hahn:2000kx} and \texttt{FeynCalc} \cite{Mertig:1990an, Shtabovenko:2016sxi, Shtabovenko:2020gxv, Shtabovenko:2023idz} packages. The numerical evaluation is performed with the \texttt{LoopTools} \cite{Hahn:1998yk} package. The largest contributions are of order $10^{-8}$ and occur around $\sqrt{M_{12}^2}\simeq 40$ GeV. As a function of $\cos\theta$, these values are reached at $-1$ and $+1$ for the left- and right-polarized amplitudes, respectively. This behavior is consistent with the findings reported in  Ref.~\cite{Thomas:2024dwd}, where Monte Carlo simulations of Higgs decays into polarized $W^\pm$ bosons including a $CP$-odd form factor were analyzed. For the case in which the $W^+$ boson is off-shell, we find a similar behavior to that shown in Fig.~\ref{Plotddif}.
 
   The pattern observed in $\cos\theta$ signals the presence of a forward-backward asymmetry ($\mathcal{A}_{FB}$), defined as
\begin{equation}
\label{FB1}
\mathcal{A}^{V}_{FB}=\frac{\int_0^1d\cos\theta\frac{ d\Gamma(H\to \overline{f}_if_jW)}{d\cos\theta dM_{12}^2}-\int_{-1}^0d\cos\theta\frac{d\Gamma(H\to \overline{f}_if_jW)}{d\cos\theta dM_{12}^2}}{\int_0^1d\cos\theta\frac{d\Gamma(H\to \overline{f}_if_jW)}{d\cos\theta dM_{12}^2}+\int_{-1}^0d\cos\theta\frac{d\Gamma(H\to \overline{f}_if_jW)}{d\cos\theta dM_{12}^2}}\text{,}\quad V=W^+\text{, }W^-,
\end{equation}
where, once again, the superscript $V$ labels the off-shell $W$ boson in the $H\to W^\ast W$ process. 
Using Eqs.~\ref{doubdifEq2}-\ref{pol0d} and carrying out the integration, we obtain
\begin{equation}
\label{FB2}
\mathcal{A}^{V}_{FB}=\frac{-\,24 \,\,
      m_W^4 \, M_{12}^2 \,
      \sqrt{(m_H^2 - m_W^2)^2 - 2(m_H^2 + m_W^2) M_{12}^2 + M_{12}^4}
}{f(m_W,m_H)}\Big({\rm Re}\big[h_1^{V}\big]\,{\rm Im}\big[h_3^{V}\big]
      - {\rm Im}\big[h_1^{V}\big]\,{\rm Re}\big[h_3^{V}\big]\Big),
\end{equation}
with the function $f(m_W,m_H)$ given by
\begin{align}
f(m_W,m_H) = &(m_H - m_W)^2 (m_H + m_W)^2 \Big\lbrace
\big({\rm Im}\big[h_2^{V}\big]^2 + {\rm Re}\big[h_2^{V}\big]^2\big) m_H^4
-2\big(-2\,{\rm Im}\big[h_1^{V}\big]\,{\rm Im}\big[h_2^{V}\big]
+{\rm Im}\big[h_2^{V}\big]^2 \nonumber\\
&
+{\rm Re}\big[h_2^{V}\big](-2\,{\rm Re}\big[h_1^{V}\big] + {\rm Re}\big[h_2^{V}\big])\big) m_H^2 m_W^2
+\big((-2\,{\rm Im}\big[h_1^{V}\big] + {\rm Im}\big[h_2^{V}\big])^2 \nonumber\\
&
+(-2\,{\rm Re}\big[h_1^{V}\big] + {\rm Re}\big[h_2^{V}\big])^2\big)m_W^4 \Big\rbrace
+ 4\Big(-\big({\rm Im}\big[h_2^{V}\big]^2+{\rm Re}\big[h_2^{V}\big]^2\big)m_H^6 
+\big\lbrace-3\,{\rm Im}\big[h_1^{V}\big]\,{\rm Im}\big[h_2^{V}\big] \nonumber\\
& 
+ {\rm Im}\big[h_2^{V}\big]^2 
+{\rm Re}\big[h_2^{V}\big]\big(-3\,{\rm Re}\big[h_1^{V}\big] + {\rm Re}\big[h_2^{V}\big]\big) 
+ 2\big({\rm Im}\big[h_3^{V}\big]^2+{\rm Re}\big[h_3^{V}\big]^2\big)\big\rbrace m_H^4 m_W^2 \nonumber\\
&
+\big\lbrace-2\,{\rm Im}\big[h_1^{V}\big]^2 - 2\,{\rm Re}\big[h_1^{V}\big]^2 
+2\,{\rm Im}\big[h_1^{V}\big]\,{\rm Im}\big[h_2^{V}\big] + {\rm Im}\big[h_2^{V}\big]^2
+2\,{\rm Re}\big[h_1^{V}\big]\,{\rm Re}\big[h_2^{V}\big] \nonumber\\
&
+ {\rm Re}\big[h_2^{V}\big]^2 
-4\big({\rm Im}\big[h_3^{V}\big]^2+{\rm Re}\big[h_3^{V}\big]^2\big)\big\rbrace m_H^2 m_W^4
+\big\lbrace10\,{\rm Im}\big[h_1^{V}\big]^2 + 10\,{\rm Re}\big[h_1^{V}\big]^2
+{\rm Im}\big[h_1^{V}\big]\,{\rm Im}\big[h_2^{V}\big] \nonumber\\
&
 - {\rm Im}\big[h_2^{V}\big]^2
+{\rm Re}\big[h_1^{V}\big]\,{\rm Re}\big[h_2^{V}\big] - {\rm Re}\big[h_2^{V}\big]^2 
+2\big({\rm Im}\big[h_3^{V}\big]^2+{\rm Re}\big[h_3^{V}\big]^2\big)\big\rbrace m_W^6 \Big) M_{12}^2\nonumber\\
&
+2\Big( 3\big({\rm Im}\big[h_2^{V}\big]^2+{\rm Re}\big[h_2^{V}\big]^2\big)m_H^4 
+2\big\lbrace 3\,{\rm Im}\big[h_1^{V}\big]\,{\rm Im}\big[h_2^{V}\big] + {\rm Im}\big[h_2^{V}\big]^2
+{\rm Re}\big[h_2^{V}\big]\big(3\,{\rm Re}\big[h_1^{V}\big] \nonumber\\
&
 + {\rm Re}\big[h_2^{V}\big]\big) 
-4\big({\rm Im}\big[h_3^{V}\big]^2+{\rm Re}\big[h_3^{V}\big]^2\big)\big\rbrace m_H^2 m_W^2
+\big\lbrace2\,{\rm Im}\big[h_1^{V}\big]^2+2\,{\rm Re}\big[h_1^{V}\big]^2 \nonumber\\
&
+2\,{\rm Im}\big[h_1^{V}\big]\,{\rm Im}\big[h_2^{V}\big] + 3\,{\rm Im}\big[h_2^{V}\big]^2
+2\,{\rm Re}\big[h_1^{V}\big]\,{\rm Re}\big[h_2^{V}\big] + 3\,{\rm Re}\big[h_2^{V}\big]^2 
-8\big({\rm Im}\big[h_3^{V}\big]^2\nonumber\\
&
+{\rm Re}\big[h_3^{V}\big]^2\big)\big\rbrace m_W^4 \Big) M_{12}^4
-4\Big( \big({\rm Im}\big[h_2^{V}\big]^2+{\rm Re}\big[h_2^{V}\big]^2\big) m_H^2
+\big({\rm Im}\big[h_2^{V}]\big({\rm Im}\big[h_1^{V}\big]+{\rm Im}\big[h_2^{V}\big]\big)\nonumber\\
&
+{\rm Re}\big[h_2^{V}]\big({\rm Re}\big[h_1^{V}\big]+{\rm Re}\big[h_2^{V}\big]\big) 
-2({\rm Im}\big[h_3^{V}\big]^2+{\rm Re}\big[h_3^{V}\big]^2)\big)m_W^2 \Big) M_{12}^6\nonumber\\
&
+\big({\rm Im}\big[h_2^{V}\big]^2+{\rm Re}\big[h_2^{V}\big]^2\big) M_{12}^8.
\end{align}

We note that the generation of a non-zero forward-backward asymmetry requires the presence of $CP$-odd and complex form factors.  In the SM, the $h_1^{W^\pm}$ and $h_3^{W^\pm}$ are induced at one-loop level, with the $CP$-odd form factor being complex. As a consequence,  Eq. \eqref{FB2} implies that the forward-backward asymmetry $\mathcal{A}_{FB}$ is non-zero. Furthermore, since $\mathcal{A}_{FB}$ depends linearly on $h_3^{V}$, the asymmetry changes sign depending on whether $W^\pm$ boson is off-shell. The structure of Eq.~\eqref{FB2} is analogous to that of the $H^\ast\to ZZ$ decay, where unpolarized angular observables are likewise proportional to the interference between the $h_1^{V}$ and $h_3^{V}$ form factors of the $HZZ$ coupling \cite{Hernandez-Juarez:2024zpk}. The connection between transverse polarizations, $CP$-odd form factors, and angular observables has been previously explored in Refs.~\cite{Barger:1993wt, Takubo:2010tc}.

\begin{figure}[!htb]
\begin{center}
\includegraphics[width=9cm]{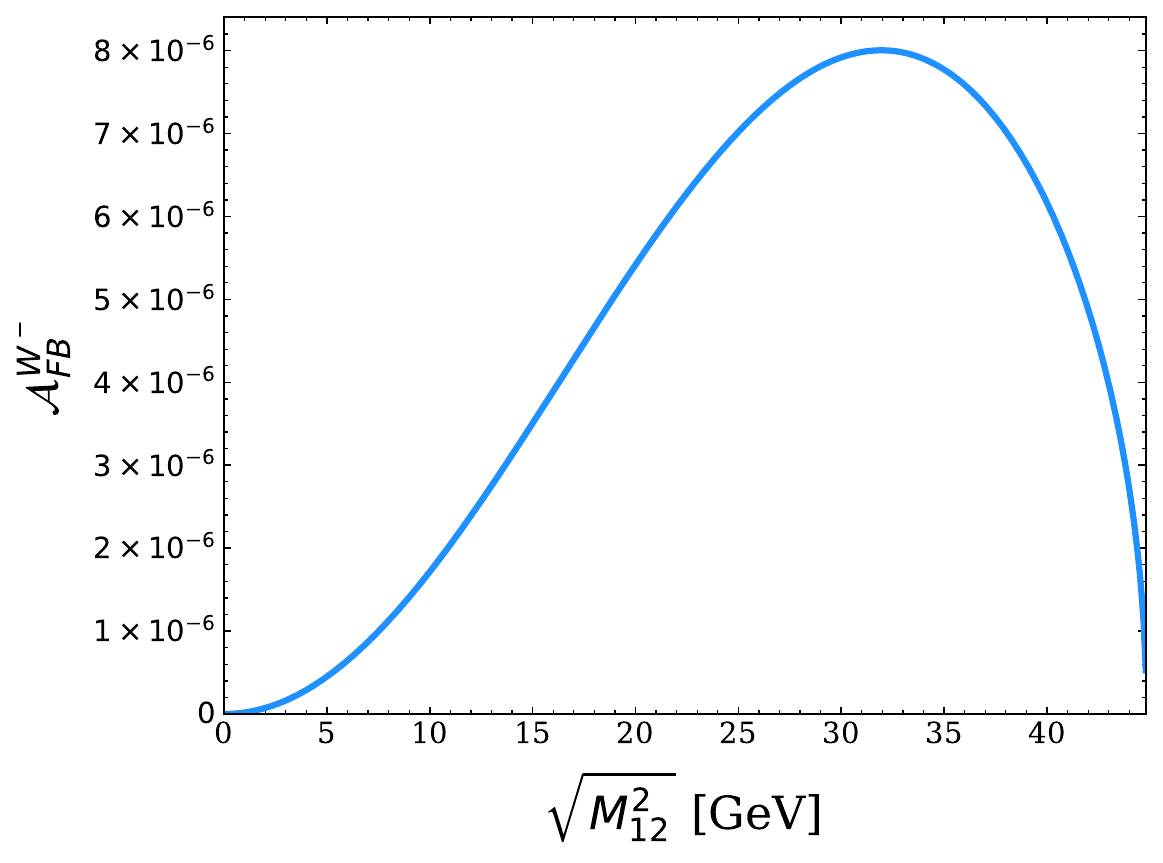}
\caption{Forward-backward asymmetry $\mathcal{A}^{W^-}_{FB}$ as a function of $\sqrt{M^2_{12}}$ within the SM. The curve changes sign when the off-shell boson corresponds to $W^{+}$.} \label{asfbplot}
\end{center}
\end{figure}

In Figure~\ref{asfbplot}, we show the behavior of $\mathcal{A}^{W^-}_{FB}$ as a function of the invariant mass $\sqrt{M_{12}^2}$. Using the SM one-loop contributions to the form factors $h_i^{W^-}$ ($i=1,2,3$), we find that $\mathcal{A}_{FB}^{W^-}$ attains values of order $10^{-6}$, with a maximum around $\sqrt{M_{12}^2}\simeq 32$ GeV. This observable is strongly suppressed at both low and high $\sqrt{M_{12}^2}$. The observation of a non-zero value of this asymmetry would point to effects of the $CP$-odd form factor within the SM that have not been fully accounted for in previous analyses.

\subsection{Left-right asymmetry }

  After integrating over $\cos\theta$, the polarized partial widths can be expressed in terms of the invariant mass of the final-state fermion pair ($M_{12}$) as
  \begin{equation}
\frac{d\Gamma^{\lambda}(H\to \overline{f} f W_\lambda)}{dM_{12}^2}= \frac{g^4 M_{12}^2 N_c \big| V_{ij}\big|^2 \sqrt{
-2 M_{12}^2 (m_H^2 + m_W^2)
+ M_{12}^4
+ (m_H^2 - m_W^2)^2
}
}{ 
3072 \pi^3 m_H^3 m_W^2(m_W^2 - M_{12}^2 )^2
}  \mathcal{M}^2_{\lambda}(M^2_{12})\text{,}\quad \lambda=L\text{, }\ R\text{, }\ 0,
\end{equation}
where $\big| V_{ij}\big|^2=1$ for leptons in the final state, and the integration range is $0\leq M_{12}^2\leq (m_H-m_W)^2$. The left- and right-polarized amplitudes are given by
\begin{align}\label{ampLR}
 \mathcal{M}_{L, R}^2=&M_{12}^2\Big( M_{12}^2 \Big({\rm Re}\big[h_3^{V}\big]{}^2 + {\rm Im}\big[h_3^{V}\big]{}^2\Big)
- 2  (m_H^2 + m_W^2) \Big({\rm Re}\big[h_3^{V}\big]{}^2 + {\rm Im}\big[h_3^{V}\big]{}^2\Big)\Big)
 \nonumber\\
  &\pm4 m_W^2 \
\sqrt{
-2 M_{12}^2 (m_H^2 + m_W^2)
+ M_{12}^4
+ (m_H^2 - m_W^2)^2
}\Big(
{\rm Re}\big[h_1^{V}\big]\, {\rm Im}\big[h_3^{V}\big]
- {\rm Im}\big[h_1^{V}\big]\, {\rm Re}\big[h_3^{V}\big]
\Big)
  \nonumber\\&+ 4m_W^4\Big({\rm Re}\big[h_1^{V}\big]{}^2 + {\rm Im}\big[h_1^{V}\big]{}^2\Big)
  \nonumber\\
  &+ (m_H^2 - m_W^2)^2 \Big({\rm Re}\big[h_3^{V}\big]{}^2 + {\rm Im}\big[h_3^{V}\big]{}^2\Big).
\end{align}
As discussed above, the longitudinally polarized component is not sensitive to the $CP$-odd form factor. For completeness, this contribution is included in Appendix~\ref{0ampl}. The calculation of polarized observables in the process $W^\ast\to W H$ has been presented in Ref.~\cite{Nakamura:2017ihk}. Summing over polarizations and taking $h_1^{V}=N_c=\big|V_{ij}\big|=1$ and $h_2^{V}=h_3^{V}=0$, we recover the tree-level unpolarized SM result for the $H\to W^*W\to \overline{f}_if_j W$ decay \cite{Pocsik:1980ta, Rizzo:1980gz, Keung:1984hn}:
\begin{equation}\label{smwidth}
    \frac{d\Gamma^{\text{SM}}}{d M^2_{12}}=\frac{g^4\sqrt{
-2 M_{12}^2 (m_H^2 + m_W^2)
+ M_{12}^4
+ (m_H^2 - m_W^2)^2
}}{3072\pi^3 m_H^3(m_W^2-M_{12}^2)^2}\Big\{ (m_H^2-m_W^2)^2+M_{12}^4-2M_{12}^2(m_H^2-5m_W^2) \Big\}.
\end{equation}

In Eq.~\eqref{ampLR}, we observe that the sign difference in the interference terms involving $h_3^{V}$ and $h_1^{V}$ persists after integration over $\cos\theta$. Two remarks are in order. First, these interference terms cancel upon summing over polarizations, so only the quadratic dependence on the $h_3^{V}$ form factor remains. As a result, in unpolarized observables the $CP$-odd effects are strongly suppressed with respect to those induced by the $CP$-conserving form factor $h_1^{V}$. Second, the impact of $h_3^{V}$ is significantly more notable in polarized quantities. For this reason, we define a left-right asymmetry $\mathcal{A}^{V}_{LR}$ for the three-body decay as
\begin{equation}
\label{LR2}
\mathcal{A}^{V}_{LR}=\frac{\frac{d\Gamma^{L}(H\to \overline{f}_if_jW_L)}{dM_{12}^2}-\frac{d\Gamma^{R}(H\to \overline{f}_if_jW_R)}{dM_{12}^2}}{\frac{d\Gamma^{L}(H\to \overline{f}_if_jW_L)}{dM_{12}^2}+\frac{d\Gamma^{R}(H\to \overline{f}_if_jW_R)}{dM_{12}^2}}\text{,}\quad V=W^+\text{, } W^-,
\end{equation}
which can be expressed analytically as 
\begin{gather}
\mathcal{A}^V_{LR} = 
\frac{
4 m_W^2 
\sqrt{
-2 M_{12}^2 (m_H^2 + m_W^2)
+ M_{12}^4
+ (m_H^2 - m_W^2)^2
}
}{G(m_H^2,m_W^2)}\Big(
{\rm Re}\big[h_1^V\big]\,{\rm Im}\big[h_3^V\big]
- {\rm Im}\big[h_1^V\big]\,{\rm Re}\big[h_3^V\big]
\Big)
\end{gather}
with the $G(m_H^2,m_W^2)$ function given as
\begin{align}
G(m_H^2,m_W^2)=&4 {\rm Im}\big[h_1^V\big]{}^2 m_W^4
+ 4 {\rm Re}\big[h_1^V\big]{}^2 m_W^4
- 2 M_{12}^2 ( {\rm Im}\big[h_3^V\big]{}^2 + {\rm Re}\big[h_3^V\big]{}^2 )(m_H^2 + m_W^2) \nonumber \\
&
+ M_{12}^4 ( {\rm Im}\big[h_3^V\big]{}^2 + {\rm Re}\big[h_3^V\big]{}^2 ) 
+ {\rm Im}\big[h_3^V\big]{}^2 (m_H^2 - m_W^2)^2
+ {\rm Re}\big[h_3^V\big]{}^2 m_H^4\nonumber
\\
&- 2 {\rm Re}\big[h_3^V\big]{}^2 m_H^2 m_W^2
+ {\rm Re}\big[h_3^V\big]{}^2 m_W^4.
\end{align}
As in the forward-backward case, non-vanishing values of $\mathcal{A}^{V}_{LR}$ require complex and $CP$-odd form factors. This observable also depends linearly on the real and imaginary parts of $h_3^{V}$. Therefore, it changes sign depending on the charge of the off-shell $W$ boson. In the SM, the left-right asymmetry $\mathcal{A}_{LR}^{V}$ is non-vanishing.

\begin{figure}[!htb]
\begin{center}
\includegraphics[width=9cm]{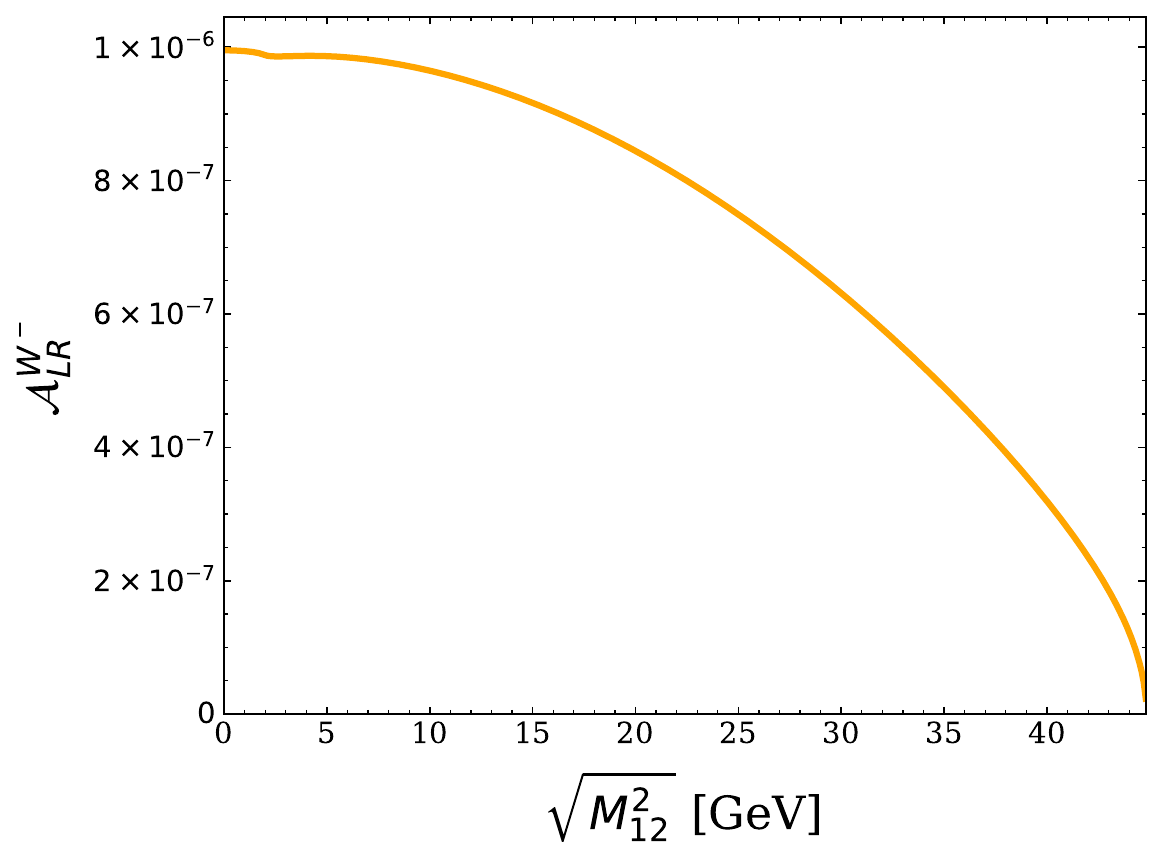}
\caption{Left-right asymmetry $\mathcal{A}^{W^-}_{LR}$ as a function of $\sqrt{M^2_{12}}$ within the SM. For an off-shell $W^{+}$ boson, the asymmetry changes sign.} \label{asLRplot}
\end{center}
\end{figure}

In Fig.~\ref{asLRplot}, we show the $\mathcal{A}_{LR}^{W^-}$ as a function of $\sqrt{M_{12}^2}$, using the one-loop SM contributions to the $h_1^{W^-}$ and $h_3^{W^-}$ form factors. Its behavior differs from that of the forward-backward asymmetry shown in Fig.~\ref{asfbplot}, as the largest values occur at low $M_{12}^2$ and are of order $10^{-6}$. At higher energies, the magnitude of $\mathcal{A}^{W^\pm}_{LR}$ decreases and reaches values of order $10^{-7}$.

As discussed in Sec.~\ref{CPsm}, the $CP$-odd form factor $h_3^H$ is not induced when both $W$ bosons are on shell. Nevertheless, for completeness, we analyze its implications in the $H^\ast\to W^- W^+$ decay in Appendix~\ref{Hoff}.

\section{Experimental prospects at the LHC}\label{prospectsSec}

At this point, one may ask whether the asymmetries discussed in this work can be observed at the LHC. For angular observables in the $H\to 2\ell 2\nu$ channel at $\sqrt{s}=13$ TeV, current sensitivity to the effects of a $CP$-odd form factor is of order $10^{-2}$ \cite{Thomas:2024dwd}. With integrated luminosities of 500-1000 fb$^{-1}$ and assuming ideal detection efficiency, sensitivities approaching $10^{-3}$ could be achievable \cite{Rao:2018abz}. These values remain three to four orders of magnitude larger than the SM prediction for the forward-backward asymmetry shown in Fig.~\ref{asfbplot}. 

For polarized observables, current sensitivities to $CP$-odd effects are of order $10^{-1}$ at $\sqrt{s}=13$ TeV \cite{Thomas:2024dwd}, i.e. about five orders of magnitude above the SM expectations in Fig.~\ref{asLRplot}. Recent developments in gauge-invariant techniques for extracting polarization information \cite{Maina:2021xpe, Javurkova:2024bwa}, together with the growing phenomenological interest in polarized gauge bosons \cite{Ballestrero:2019qoy, Maina:2020rgd, Denner:2020bcz, Denner:2021csi, Maina:2021xpe, Denner:2023ehn, Javurkova:2024bwa, Grossi:2024jae, Thomas:2024dwd, Denner:2024tlu, Colyer:2025ehv, Haisch:2025jqr, Pelliccioli:2025com}, suggest that substantial progress can be expected in the near future. The physics of polarized gauge bosons has been implemented in event generators such as \texttt{MadGraph5\_aMC@NLO} \cite{BuarqueFranzosi:2019boy} and \texttt{SHERPA} \cite{Hoppe:2023uux}, thereby enabling more accurate theoretical predictions for polarized observables. On the experimental side, measurements of $Z$ and $W^\pm$ polarization are already being carried out by ATLAS \cite{ATLAS:2012au, ATLAS:2016rnf, ATLAS:2016fbc, ATLAS:2019bsc, CMS:2020ezf, ATLAS:2022oge, ATLAS:2022rms, ATLAS:2023lsr, ATLAS:2023zrv, ATLAS:2025wuw}, CMS \cite{CMS:2011kaj, CMS:2015cyj, CMS:2016asd, CMS:2020ezf, CMS:2020etf}, and LHCb \cite{LHCb:2022tbc} collaborations. Improvements in sensitivity, both for angular and polarized observables, may also arise from the use of deep neural networks, such as those implemented by ATLAS and CMS collaborations. The asymmetries analyzed in this work provide a stringent test of the SM predictions. Their experimental confirmation would signal the presence of $CP$-odd effects that have remained largely unexplored in previous studies.

\section{Conclusions}\label{concSec}

In this work, we have carried out the first one-loop evaluation of the $CP$-odd contribution to the $HWW$ vertex within the SM. We find that the corresponding form factor $h_3^{W^\pm}$ is complex and dominated by its imaginary part. The largest contribution arises from the $\overline{t}bW^\pm$ coupling and is of order $10^{-5}$, well below the current experimental sensitivity of order $10^{1}$. We also show that $h_3^{W^\pm}$ is generated only when the two $W$ bosons have different virtualities ($p_1^2\neq p_2^2$).

The form factor $h_3^{W^\pm}$ is not $CP$-violating, as it is not proportional to the Jarlskog invariant. Nevertheless, its contribution constitutes an irreducible SM background in searches for $CP$ violation in the $HWW$ coupling. We therefore investigated the effects of $h_3^{W^\pm}$ in the decay $H\to \overline{f}_if_j W^\pm$.  Furthermore, we defined the forward-backward ($\mathcal{A}^{W^\pm}_{FB}$) and left-right ($\mathcal{A}^{W^\pm}_{LR}$) asymmetries. These observables are non-vanishing only in the presence of complex and $CP$-odd couplings. In the SM, $\mathcal{A}^{W^\pm}_{FB}$ and $\mathcal{A}^{W^\pm}_{LR}$ are non-zero and reach magnitudes of order $10^{-6}$ and $10^{-7}$, respectively. These values are far beyond the reach of current experiments. Nevertheless, recent developments in methods for accessing polarized observables may significantly enhance experimental sensitivity. The observation of these asymmetries would be evidence of a previously missed  $CP$-odd form factor within the SM.

\begin{acknowledgments}

The authors thank the referee for insightful comments and suggestions that significantly improved the quality and clarity of this work. We thank I. Garc\'ia-M\'arquez for assistance in the preparation of Fig. \ref{plano}.
This work  was supported by  "Estancias Posdoctorales por M\'exico (SECIHTI)". We also acknowledge support from  Sistema Nacional de Investigadores (Mexico). 
\end{acknowledgments}


\appendix

\section{$\text{B}_0$ scalar functions}\label{b0ap}

The scalar Passarino--Veltman $\text{B}_0$ functions appearing in Eq.~\eqref{h3} are given by
\begin{equation}
\text{B}_0(p^2,m_i^2,m_j^2)=\Delta-\int^1_0 dx \ln\Bigg[\frac{-x(1-x)p^2+x\, m_j^2+(1-x)m_i^2}{\mu^2}\Bigg],
\end{equation}
with 
\begin{equation}
\Delta=\frac{2}{\epsilon}-\gamma+\ln(4\pi),
\end{equation}
where the ultraviolet divergence of the $\text{B}_0$ functions appears as a pole in the limit $\epsilon\to0$.

\section{Kinematics}
\subsection{$H\to \overline{f}(p_1)f(p_2)W(p_3)$}\label{kin1}
 The kinematics of the $H(q)$ and $W(p_3)$ bosons in the rest frame of the Higgs boson can be described by the following relations.
\begin{align}
    & q^\mu=\big(m_H,\ 0  \big),  
    &  p_{3}^\mu=\bigg(\frac{m_H^2+m_W^2-M_{12}^2}{2m_H}, \vec{p}_3\bigg),
\end{align}
  where the magnitude of the three-momentum $\vec{p}_3$ is 
\begin{equation}
\|\vec{p}_3\|=\frac{\sqrt{(m_H^2+m_W^2-M_{12}^2)^2-4 m_H^2m_W^2}}{2m_H}.
\end{equation}
\begin{figure}[H]
\begin{center}
\includegraphics[width=8cm]{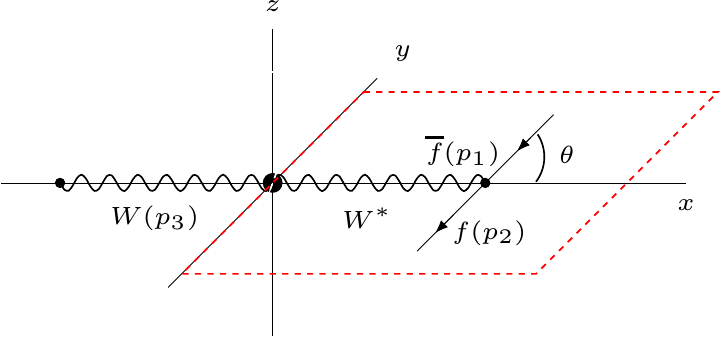}
\caption{Kinematics for the $H\to\overline{f}_if_jW$ decay.} \label{plano}
\end{center}
\end{figure}

We consider that in this frame the $W(p_3)$ gauge boson moves along the negative $x$ axis, so their polarization vectors can be written as
\begin{align}
  & \epsilon^\mu_{3}(0)=\frac{1}{m_W}\Big( \| \vec{p}_3 \|, -\frac{m_H^2+m_W^2-M_{12}^2}{2m_H} ,0  , 0  \Big), \\
    &  \epsilon^\mu_{3}(R/L)=\frac{1}{\sqrt{2}}\Big(0 , 0 ,  i  , \pm1  \Big). \label{polRLvecH}
\end{align}
The fermion momenta in their center of mass system are given by
\begin{align}
    &p_1^\mu=\frac{M_{12}}{2}(1,\hat{n}_1),\\
    &p_2^\mu=\frac{M_{12}}{2}(1,-\hat{n}_1),
\end{align}
with the unit vector
\begin{equation}
    \hat{n}_1=(\cos{\theta},\sin{\theta},0).
\end{equation}
Here, $\theta$ is the angle between the $\hat{x}$ and $\vec{p}_1$ vectors in the $\overline{f}(p_1)f(p_2)$ center of mass as shown in Fig. \ref{plano}. 
We define the $M_{ij}^2$ invariant masses as follows:
\begin{align}
    &M_{ij}^2=(p_i+p_j)^2.\label{Minv}
\end{align}
Note that for massless fermions the $M_{ij}^2$ masses fulfill the relation 
\begin{equation}
    m_H^2 = M_{31}^2 + M_{12}^2 + M_{23}^2 - m_W^2
\end{equation}
The fermion momenta can be boosted to the Higgs rest frame through the following velocity
\begin{equation}
    v=\frac{\sqrt{(m_H^2+m_W^2-M_{12}^2)^2-4 m_H^2m_W^2}}{m_H^2+m_W^2-M_{12}^2}.
\end{equation}
In this frame, the relevant invariant is
\begin{equation}
    p_2\cdot p_3=\frac{1}{4} \left( 
    -\cos\theta \sqrt{ 
        -2 M_{12}^2 (m_H^2 + m_W^2) + M_{12}^4 + (m_H^2 - m_W^2)^2 
    } 
    - M_{12}^2 + m_H^2 - m_W^2 
\right).\label{scp}
\end{equation}
From Eq. \eqref{Minv} and \eqref{scp}, we find that
\begin{equation}\label{dif}
    d(M_{23}^2) = -\frac{1}{2} \sqrt{(m_H^2 - m_W^2)^2 - 2(m_H^2 + m_W^2) M_{12}^2 + M_{12}^4} \; d(\cos\theta),
\end{equation}
where the range of the angle $\theta$ is
\begin{equation}
    -1\leq \cos{\theta} \leq 1.
\end{equation}

\subsection{$H^\ast\to WW$}
 The kinematics of the  $H^\ast\rightarrow WW$ decay in the Higgs boson rest frame can be described by the following relations
\begin{align}
    & q^\mu=\big(Q,\ 0  \big),  \\
    &  p_{1,2}^\mu=\big(Q/2,\pm \vec{p}\big),
\end{align}
  where the magnitude of the three-momentum $\vec{p}$ is 
\begin{equation}
\|\vec{p}\|=\frac{\sqrt{Q^2-4 m_W^2}}{2}.
\end{equation}
We consider that in this frame the $W$ gauge bosons move along the $x$ axis, so their polarization vectors can be written as
\begin{align}
  & \epsilon^\mu_{1,2}(0)=\frac{1}{m_W}\Big( \| \vec{p} \|, \pm \frac{\sqrt{Q^2}}{2} ,0  , 0  \Big), \\
    &  \epsilon^\mu_{1}(R/L)=\frac{1}{\sqrt{2}}\Big(0 , 0 , - i  , \pm1  \Big). \label{polRLvec}
\end{align}

\section{Longitudinally polarized amplitude}\label{0ampl}

We present the squared amplitude for the 0 polarization corresponding $\frac{d\Gamma^{\lambda}(H\to \overline{f} f W_\lambda)}{dM_{12}^2}$, which is given as follows:

\begin{align}\label{0polamp}
  \mathcal{M}^2_0=& 
     m_W^4 \Big(
 2 M_{12}^4 \Big( 
2 {\rm Im}\big[h_1^{V}\big]{}^2
+ 2 {\rm Im}\big[h_1^{V}\big]{\rm Im}\big[h_2^{V}\big]
+ 2 {\rm Re}\big[h_1^{V}\big]{}^2
+ 2 {\rm Re}\big[h_1^{V}\big]{\rm Re}\big[h_2^{V}\big]
+ 3 \big({\rm Im}\big[h_2^{V}\big]{}^2 + {\rm Re}\big[h_2^{V}\big]{}^2\big)
\Big) \nonumber\\
& + 2 m_H^2 m_W^2 \Big(
3 {\rm Im}\big[h_1^{V}\big]{\rm Im}\big[h_2^{V}\big]
+ {\rm Re}\big[h_2^{V}\big](3 {\rm Re}\big[h_1^{V}\big] + {\rm Re}\big[h_2^{V}\big])
+ {\rm Im}\big[h_2^{V}\big]{}^2
\Big) \nonumber
\\ &+ 3 m_H^4 \big({\rm Im}\big[h_2^{V}\big]{}^2 + {\rm Re}\big[h_2^{V}\big]{}^2 \big)
\Big) \nonumber\\
& - 4 M_{12}^6 \Big(
m_W^2 \big(
{\rm Im}\big[h_2^{V}\big]({\rm Im}\big[h_1^{V}\big] + {\rm Im}\big[h_2^{V}\big])
+ {\rm Re}\big[h_2^{V}\big]({\rm Re}\big[h_1^{V}\big] + {\rm Re}\big[h_2^{V}\big])
\big) 
+ m_H^2 \big({\rm Im}\big[h_2^{V}\big]{}^2 + {\rm Re}\big[h_2^{V}\big]{}^2 \big)
\Big) \nonumber\\
& - 4 M_{12}^2 (m_H - m_W)(m_H + m_W) \Big(
3 m_H^2 m_W^2 \big({\rm Im}\big[h_1^{V}\big]{\rm Im}\big[h_2^{V}\big]
+ {\rm Re}\big[h_1^{V}\big]{\rm Re}\big[h_2^{V}\big]\big) \nonumber\\
& + m_W^4 \big(
(2 {\rm Im}\big[h_1^{V}\big] - {\rm Im}\big[h_2^{V}\big])
({\rm Im}\big[h_1^{V}\big] + {\rm Im}\big[h_2^{V}\big])
+ (2 {\rm Re}\big[h_1^{V}\big] - {\rm Re}\big[h_2^{V}\big])
({\rm Re}\big[h_1^{V}\big] + {\rm Re}\big[h_2^{V}\big])
\big) \nonumber\\
& + m_H^4 \big({\rm Im}\big[h_2^{V}\big]{}^2 + {\rm Re}\big[h_2^{V}\big]{}^2\big)
\Big) \nonumber\\
& + (m_H - m_W)^2 (m_H + m_W)^2 \Big(
-2 m_H^2 m_W^2 \big(
-2 {\rm Im}\big[h_1^{V}\big]{\rm Im}\big[h_2^{V}\big]
+ {\rm Re}\big[h_2^{V}\big]({\rm Re}\big[h_2^{V}\big] - 2 {\rm Re}\big[h_1^{V}\big])
+ {\rm Im}\big[h_2^{V}\big]{}^2
\Big) \nonumber\\
& + m_W^4 \Big(
({\rm Im}\big[h_2^{V}\big] - 2 {\rm Im}\big[h_1^{V}\big])^2
+ ({\rm Re}\big[h_2^{V}\big] - 2 {\rm Re}\big[h_1^{V}\big])^2
\Big)
+ m_H^4 \big({\rm Im}\big[h_2^{V}\big]{}^2 + {\rm Re}\big[h_2^{V}\big]{}^2 \big)
\Big) \nonumber\\
& + M_{12}^8 \big({\rm Im}\big[h_2^{V}\big]{}^2 + {\rm Re}\big[h_2^{V}\big]{}^2\big).
\end{align}
It is observed that $\mathcal{M}_0^2$ is not in terms of the $CP$-odd form factor $h_3^{V}$.

\section{$H^\ast\to W^-W^+$ case}\label{Hoff}

 We present the left-right asymmetry for the $H^\ast\to W^-W^+$ process.  The polarized partial width is defined as
\begin{equation}\label{gammaHZZp}
\Gamma^{\lambda \lambda}(H^\ast\rightarrow W^-_{\lambda} W^+_{\lambda})=\frac{g^2  \sqrt{Q^2-4  m_W^2}}{128 \pi m_W^2  Q^2 }\mathcal{M}^2_{\lambda \lambda},
\end{equation}
where we take identical transverse polarizations for both bosons. The transversely polarized amplitudes are given by
\begin{align}
\label{MLL}
 \mathcal{M}^2_{LL}=&Q^2 \left(Q^2-4
   m_W^2\right) \left({\rm Re}\big[h_3^H\big]{}^2+{\rm Im}\big[h_3^H\big]{}^2\right)+4 m_W^4
   \Big({\rm Re}\big[h_1^H\big]{}^2+{\rm Im}\big[h_1^H\big]{}^2\Big)\nonumber\\
   & +4  m_W^2  \sqrt{Q^4-4 Q^2 m_W^2} \Big({\rm Re}\big[h_1^H\big]{\rm Im}\big[h_3^H\big] - {\rm Im}\big[h_1^H\big] {\rm Re}\big[h_3^H\big] \Big),
\end{align}
\begin{align}
\label{MRR}
 \mathcal{M}^2_{RR}=&Q^2 \left(Q^2-4
   m_W^2\right) \left({\rm Re}\big[h_3^H\big]{}^2+{\rm Im}\big[h_3^H\big]{}^2\right)+4 m_W^4
   \Big({\rm Re}\big[h_1^H\big]{}^2+{\rm Im}\big[h_1^H\big]{}^2\Big)\nonumber\\
   &-4  m_W^2  \sqrt{Q^4-4 Q^2 m_W^2} \Big({\rm Re}\big[h_1^H\big]{\rm Im}\big[h_3^H\big] - {\rm Im}\big[h_1^H\big] {\rm Re}\big[h_3^H\big] \Big).
   \end{align}
   The longitudinal contribution is omitted because it is not sensitive to the $CP$-odd form factor \cite{Hernandez-Juarez:2023dor}. The opposite sign of the interference terms in Eqs.~\eqref{MLL}-\eqref{MRR} motivates the definition of the left-right asymmetry $\mathcal{A}^H_{LR}$,
 \begin{equation}
\label{LR1}
\mathcal{A}^H_{LR}=\frac{\Gamma^{LL}(H^\ast\to W^-_LW^+_L)-\Gamma^{RR}(H^\ast\to W^-_RW^+_R)}{\Gamma^{LL}(H^\ast\to W^-_LW^+_L)+\Gamma^{RR}(H^\ast\to W^-_RW^+_R)}.
\end{equation}
 In terms of the $h_i^H$ form factors, the $\mathcal{A}^H_{LR}$ asymmetry can be expressed as follows
\begin{align}
\label{ALRe}
\mathcal{A}^H_{LR}= \frac{4  m_W^2 Q\sqrt{Q^2-4 m_W^2} \left({\rm Re}\big[h_1^H\big]{\rm Im}\big[h_3^H\big]
   -{\rm Re}\big[h_3^H\big]{\rm Im}\big[h_1^H\big] \right)}{Q^2
   \left(Q^2-4 m_W^2\right)
   \left({\rm Re}\big[h_3^H\big]{}^2+{\rm Im}\big[h_3^H\big]{}^2\right)+4 m_W^4\big(
   {\rm Im}\big[h_1^H\big]{}^2+ {\rm Re}\big[h_1^H\big]{}^2\big)}.
   \end{align}
As in the cases of $\mathcal{A}_{FB}^{W^\pm}$ and $\mathcal{A}_{LR}^{W^\pm}$, complex and $CP$-odd form factors are required to generate a non-zero $\mathcal{A}_{LR}^{H}$. Within the SM, this asymmetry vanishes. However, it can arise in scenarios beyond the SM, such as through loops of heavy Majorana neutrinos \cite{Ilakovac:1993pt}.  This left-right asymmetry was first analyzed in Refs. \cite{Ilakovac:1993pt, Chang:1992tu}. An analogous observable for the polarized process $H^\ast\to ZZ$ has been investigated in Refs.~\cite{Hernandez-Juarez:2023dor,Hernandez-Juarez:2024zpk,Hernandez-Juarez:2025nzh}.

\bibliography{Biblio}

\end{document}